\documentclass[aps,prd,twocolumn,superscriptaddress,preprintnumbers,nofootinbib,longbibliography,floatfix]{revtex4-1}

\usepackage{graphicx}
\usepackage{dcolumn}
\usepackage{bm}
\usepackage{enumitem}
\usepackage{algorithm}
\usepackage{algpseudocode}
\usepackage[dvipsnames]{xcolor}

\usepackage[caption=false]{subfig}

\usepackage[colorlinks=true
,urlcolor=blue
,anchorcolor=blue
,citecolor=blue
,filecolor=blue
,linkcolor=blue
,menucolor=blue
,pagecolor=blue
]{hyperref}
\usepackage[all]{hypcap}
\usepackage{braket}
\usepackage{amsmath,amsfonts}
\usepackage{comment}
\usepackage{cleveref}

\DeclareMathOperator*{\argmax}{argmax}

\DeclareRobustCommand{\Sec}[1]{Sec.~\ref{sec:#1}}

\DeclareRobustCommand{\Fig}[1]{Fig.~\ref{fig:#1}}

\DeclareRobustCommand{\Eq}[1]{Eq.~(\ref{eq:#1})}

\DeclareRobustCommand{\Ref}[1]{Ref.~\cite{#1}}
\DeclareRobustCommand{\Refs}[1]{Refs.~\cite{#1}}

\makeatletter
\def\l@subsubsection#1#2{}
\makeatother

\begin{document}

%
%

\preprint{MIT-CTP 5428}

\title{Quantum Annealing for Jet Clustering with Thrust}
\thanks{This manuscript has been authored by UT-Battelle, LLC, under Contract No.~DE-AC0500OR22725 with the U.S. Department of Energy. The United States Government retains and the publisher, by accepting the article for publication, acknowledges that the United States Government retains a non-exclusive, paid-up, irrevocable, world-wide license to publish or reproduce the published form of this manuscript, or allow others to do so, for the United States Government purposes. The Department of Energy will provide public access to these results of federally sponsored research in accordance with the DOE Public Access Plan.}%

\author{Andrea Delgado}
\email{delgadoa@ornl.gov}
\affiliation{Physics Division, Oak Ridge National Laboratory, Oak Ridge, TN 37830}

\author{Jesse Thaler}
\email{jthaler@mit.edu}
\affiliation{Center for Theoretical Physics, Massachusetts Institute of Technology, Cambridge, MA 02139}

\begin{abstract}
Quantum computing holds the promise of substantially speeding up computationally expensive tasks, such as solving optimization problems over a large number of elements.
In high-energy collider physics, quantum-assisted algorithms might accelerate the clustering of particles into jets.
In this study, we benchmark quantum annealing strategies for jet clustering based on optimizing a quantity called ``thrust'' in electron-positron collision events.
We find that quantum annealing yields similar performance to exact classical approaches and classical heuristics, but only after tuning the annealing parameters. 
Without tuning, comparable performance can be obtained through a hybrid quantum/classical approach.
\end{abstract}

\maketitle

{\tableofcontents}

%
%

\section{Introduction}
\label{sec:intro}

Quantum computers aim to harness the phenomena of quantum mechanics to deliver considerable leaps in processing power.
In this way, quantum-assisted algorithms might provide a solution to the increasingly challenging and soon intractable problem of analyzing and simulating the interaction of particles in high-energy physics experiments such as the Large Hadron Collider (LHC)~\cite{Alves:2017she}.
In particular, computationally expensive tasks such as solving optimization problems over many elements might be sped up using quantum computers.
One example of such an application in high-energy physics is that of clustering particles into jets.

In particle physics, a jet is a collection of particles collimated into a roughly cone-shaped region.
Jets arise from the fragmentation of quarks and gluons that are produced in high-energy collisions.
Because of the confining properties of quantum chromodynamics (QCD), quarks and gluons cannot be detected in isolation since they carry color charge.
Therefore, the process of jet fragmentation yields sprays of color-neutral particles that can be experimentally measured in particle detectors.
To estimate the kinematics of the quark or gluon that initiated the jet, one typically uses jet clustering algorithms to combine the observed particles into a collective jet object for further study; see \Ref{Salam:2010nqg} for a review.

In electron-positron collisions, the dominant event topology involves two back-to-back jets from the fragmentation of a quark and an anti-quark.
This motivates partitioning the event into two hemisphere jets, which can be accomplished using event shapes~\cite{PhysRevLett.35.1609}.
One popular but computationally expensive event shape is thrust~\cite{BRANDT196457,PhysRevLett.39.1587}, which involves finding the hemisphere partition that maximizes their summed three-momenta.
For an event with $N$ particles, finding the thrust optimum has a naive runtime of $O(N^3)$~\cite{YAMAMOTO1983597}, though using a trick introduced in \Ref{Salam_2007}, it is possible to improve this to $O(N^2 \log N)$~\cite{PhysRevD.101.094015}.
On a universal quantum computer, thrust can be computed in $O(N^2)$~\cite{PhysRevD.101.094015} using a strategy based on Grover search.
Alternatively, \Ref{PhysRevD.101.094015} showed how thrust can be phrased as a quadratic unconstrained binary optimization (QUBO) problem, suitable for quantum annealing.
See \Refs{Pires:2020urc,Pires:2021fka,Kim:2021wrr,deLejarza:2022bwc} for other proposed quantum algorithms for jet clustering, and \Ref{Delgado:2022tpc} for an extensive review of quantum algorithms for collider data analysis.

In this paper, we benchmark the performance of quantum annealing for hemisphere jet clustering with thrust.
Thrust is a particularly interesting optimization problem for testing quantum algorithms because it has multiple equivalent exact formulations and well-studied approximations.
For quantum annealing, we test a thrust implementation on the Advantage QPU from D-Wave.
We compare quantum annealing to its classical counterpart and also study a classical heuristic based on iterative optimization.
After tuning the annealing parameters, we find that quantum annealing yields good performance.
Even without tuning, we can combine quantum annealing with classical iterative updates to reach the same performance with less computational overhead.
This suggests the importance of further research on hybrid quantum/classical algorithms for high-energy physics.

The remainder of this paper is organized as follows.
In \Sec{review}, we review the formulation of thrust as a jet clustering algorithm and explain how it can be solved using quantum annealing.
In \Sec{quantumannealing}, we discuss the details of quantum annealing and the specifics of the D-Wave quantum processor device used for out study.
In \Sec{results}, we present the results obtained when attempting to solve the thrust-inspired quantum annealing algorithm proposed in \Ref{PhysRevD.101.094015}.
Our conclusions are provided in \Sec{conclusion}.

\section{Review of Jet Clustering with Thrust}
\label{sec:review}

Event shapes~\cite{PhysRevLett.35.1609} are among the most extensively studied observables to characterize jet-like behavior at colliders.
Their conceptual simplicity, combined with their sensitivity to a range of QCD radiation features, makes them interesting observables.
Event shapes have been studied in hadronic final states of electron-positron ($e^{+}e^{-}$) and deep inelastic scattering (DIS) collisions \cite{Dasgupta2004,2006MESV,Mele2008}. 
For our purposes, event shapes can often be interpreted as partitioning an event into jets, which we review in this section.

Our study focuses on electron-positron collisions, where the cross section is dominated by the process $e^{+}e^{-}\rightarrow\gamma^{*}/Z^{0}\rightarrow q\bar{q}$.
These events consist of a quark recoiling against an anti-quark with equal and opposite momentum at lowest order.
QCD radiation and hadronization produce deviations from this back-to-back structure, measured and quantified using event shapes.

Global event shapes like thrust~\cite{BRANDT196457,PhysRevLett.39.1587} and sphericity~\cite{PhysRevD.1.1416} are computed based on all observed final-state particles.
Each collider event is characterized by a set of particle four-momenta $p_{i} = (E_i, \vec{p}_{i})$, with the index $i=1,2, \ldots ,N$ running over all particles in the event. 
Following the event-shape literature, we typically drop the energy information $E_i$ and restrict our attention to the three-momenta $\vec{p}_{i} = (p_{i}^{x}, p_{i}^{y}, p_{i}^{z})$.
Throughout this discussion, the particle kinematics are computed in the center-of-mass frame of the collision, such that
\begin{equation}
\label{eq:center_of_mass_frame}
    \sum_{i = 1}^N \vec{p}_{i} = 0.
\end{equation}

\subsection{Thrust as a Partition Problem}
\label{sec:thrust_definition}

Thrust is arguably the best-studied event shape at electron-positron colliders.
It has been measured with high precision, especially at $Z^{0}$ peak, and compared to precision QCD calculations \cite{Abbiendi:2004qz,Heister:2003aj}.
Due to its ubiquity, it is a useful benchmark for the study of quantum algorithms for colliders.

There are multiple equivalent definitions of thrust, and we point the reader to \Ref{PhysRevD.101.094015} for a detailed discussion.%
\footnote{More recently, \Ref{Komiske:2020qhg} showed how to phrase thrust in the language of optimal transport theory.}
Originally, thrust was defined as an axis-finding problem~\cite{BRANDT196457,PhysRevLett.39.1587}, where one tries to find the unit three-vector $\hat{n}$ that maximizes the quantity
\begin{equation}
\label{eq:thrust_as_axis}
    T(\hat{n}) = \frac{\sum_{i = 1}^N|\hat{n}\cdot \vec{p_{i}}|}{\sum_{i = 1}^N |\vec{p}_{i}|}.
\end{equation}
Thrust itself is determined by
\begin{equation}
T = \max_{\hat{n}} T(\hat{n}),
\end{equation}
with corresponding thrust axis
\begin{equation}
\hat{n}_T = \argmax_{\hat{n}} T(\hat{n}).
\end{equation}
Thrust takes on a maximum value of 1 for two back-to-back particles and a minimum of $1/2$ for a perfectly isotropic event configuration.
For the results in \Sec{results}, we report the value of $1-T \in [0, 0.5]$, to emphasize the behavior in the dijet ($1-T \to 0$) limit.

In the context of jet clustering, it is more convenient to phrase thrust as a partitioning problem~\cite{Yamamoto:1984fd}.
Let $x_i$ equal 1 if particle $i$ is inside of a jet and 0 otherwise.
From a given partition, we can compute the quantity
\begin{equation}
\label{eq:thrust_as_partition}
T(\{x_i\}) = 2 \frac{\Big| \sum_{i = 1}^N x_i \, \vec{p}_{i} \Big|}{\sum_{i = 1}^N |\vec{p}_{i}|},
\end{equation}
where the factor of 2 accounts for the fact that the particles inside and outside of the jet have equal and opposite momenta in the center-of-mass frame.
Thrust turns out to be equivalent to 
\begin{equation}
T = \max_{\{x_i\}} T(\{x_i\}).
\end{equation}
Letting $\{\tilde{x}_i\}$ be the partition that maximizes the above expression, the jet three-momentum is
\begin{equation}
\vec{P} = \sum_i \tilde{x}_i \, \vec{p}_{i}.
\end{equation}
By momentum conservation, the total three-momentum of the particles outside of the jet is $-\vec{P}$, which means that one can replace $\tilde{x}_i \to 1 - \tilde{x}_i$ with no change to the thrust value.

To understand the equivalence of the axis-based and partition-based formulations, note that the jet three-vector $\vec{P}$ and the thrust axis $\hat{n}_T$ are parallel.
Particles are inside the jet if $\hat{n}_T \cdot \vec{p_{i}} > 0$ and outside otherwise, which means that the thrust axis effectively partitions the event into two hemispheres separated by a plane perpendicular to the thrust axis.
(For a finite number of particles, $\hat{n}_T \cdot \vec{p_{i}}$ can never equal zero.)
Often, one talks about the particles with $x_i = 1$ as being in the left hemisphere while those with $x_i = 0$ being in the right hemisphere.

To compute thrust exactly, we use the algorithm in \Ref{YAMAMOTO1983597}, which is implemented in the event generator \textsc{Pythia} 8~\cite{Sj_strand_2008}.
This algorithm finds the exact value of thrust by searching over all possible partitions of the event into two hemispheres.
Because two three-vectors span a plane, there are $O(N^2)$ partitions to check.
Computing \Eq{thrust_as_partition} for a fixed partition takes $O(N)$, yielding an $O(N^3)$ algorithm.

\subsection{Iterative Approximation}
\label{sec:iterative}

A classical strategy to approximate thrust is by using an expectation-maximization-style algorithm.
(This approach is used in the thrust implementation of \textsc{Pythia} 6~\cite{Sjostrand:2006za}.)
Starting from a seed partition, one can compute a seed jet three-momentum $\vec{P}_{\rm seed}$.
Then, one can find an updated partition in $O(N)$ by splitting the event into hemispheres separated by the plane perpendicular to $\vec{P}_{\rm seed}$.
This process can be iterated and will converge in a finite number of steps.
While this method is not guaranteed to find the true thrust optimum, it is a computationally efficient method to find the local minimum.

\subsection{Seed Axis via Sphericity}
\label{sec:sphericity}

One way to choose a sensible seed axis for the iterative approximation above is to compute another event shape called sphericity.
As a QCD event shape, sphericity has no specific disadvantages over thrust, though as discussed in \Ref{Cesarotti:2020hwb}, it is not a true measure of isotropy.

The generalized sphericity tensor is defined as
\begin{equation}
    S^{ab} = \frac{\sum_{i} p_i^a p_i^b |\vec{p}_i|^{r-2}}{\sum_i |p_i|^2},
    \label{eq:sph}
\end{equation}
where $a,b = 1,2,3$ corresponds to the $x,y,z$ components of the particle three-momenta.
The original sphericity tensor in \Ref{PhysRevD.1.1416} uses $r = 2$, but we focus on the linearized sphericity tensor with $r = 1$ since it is infrared and collinear safe.

By diagonalizing $S^{ab}$, one finds three eigenvalues $\lambda_{1}\geq \lambda_{2} \geq \lambda_{3}$, with $\lambda_{1}+\lambda_{2}+\lambda_{3} = 1$.
The sphericity of the event is then defined as
\begin{equation}
    S = \frac{3}{2}(\lambda_{2} + \lambda_{3}),
\end{equation}
such that $S = 0$ for two back-to-back particles and $S = 1$ for configurations with equal eigenvalues.

The sphericity axis is the eigenvector associated with the eigenvalue $\lambda_{1}$.
The sphericity axis is similar to, but not identical to, the thrust axis, and is therefore a useful seed for the iterative approach.

\subsection{Quantum Annealing for Thrust}

The first quantum-assisted algorithms for jet clustering were proposed in \Ref{PhysRevD.101.094015}, based on thrust optimization.
In the context of quantum annealing, thrust can be rephrased as a quadratic unconstrained binary optimization (QUBO) problem with objective function
\begin{equation}
    \label{eq:qubo}
    O_{\rm QUBO}(\{x_i\}) = \sum_{i,j=1}^{N} \vec{p_{i}}\cdot\vec{p_{j}} \, x_{i} \, x_{j},
\end{equation}
where once again, each $x_i$ takes the value 0 or 1.
To relate this to thrust, note that
\begin{equation}
\label{eq:thrust_as_qubo}
    O_{\rm QUBO}(\{x_i\}) = \Big( \sum_{i = 1}^N |\vec{p}_{i}| \Big)^2 \, T(\{x_i\})^2,
\end{equation}
where $T(\{x_i\})$ is defined in \Eq{thrust_as_partition}.
Because the term in parentheses is independent of the partition, finding the maximum of $O_{\rm QUBO}$ is the same as optimizing for thrust.
This objective function will be the basis for our quantum annealing studies.

Through we do not study it here, \Eq{qubo} is part of a one-parameter family of QUBO-based jet algorithms~\cite{PhysRevD.101.094015}:
\begin{align}
\label{eq:qubo_singlecone}
    O_{\rm QUBO}(\{x_i\}) &= \sum_{i,j=1}^{N} Q_{ij} \, x_{i} \, x_{j},\\
    Q_{ij} &= \frac{\vec{p_{i}}\cdot\vec{p_{j}} - E_{i}\, E_{j}\cos R}{1 - \cos R},
\end{align}
where $Q_{ij}$ is the QUBO matrix.
This expression is a variant of the \textsc{SISCone} algorithm~\cite{Salam_2007}, and dubbed as \textsc{SingleCone} in \Ref{PhysRevD.101.094015}.
This variant aims to cluster particles into a single jet with characteristic radius $R$.
Taking $R$ to $\pi/2$ reduces to the thrust (squared) problem in \Eq{thrust_as_qubo}.

\section{Quantum Annealing}
\label{sec:quantumannealing}

Two quantum strategies for computing thrust were presented in \Ref{PhysRevD.101.094015}, one based on universal quantum computing~\cite{Benioff1980,Feynman1982, 1985QT} and one based on quantum annealing~\cite{Johnson2011,Harris_2010,ApolloniCesa, APOLLONI1989233, PhysRevE.58.5355,FINNILA1994343}.
These two paradigms for quantum computing are very different in their modes of operation, currently available system sizes, and types of problems they can solve.
In particular, quantum annealing excels at solving optimization problems such as those formulated in QUBO form.
For this study, we focus exclusively on quantum annealing for thrust. 

\subsection{D-Wave Advantage QPU}
\label{sec:advantage_qpu}

D-Wave offers a variety of commercially available quantum annealing devices~\cite{Dwave-QPU}, whose operation is based on the adiabatic theorem~\cite{Born1928}.
Starting from an initial Hamiltonian whose ground state is known, the system is slowly evolved to the problem Hamiltonian of interest.
If the system is evolved sufficiently slowly, then the adiabatic theorem ensures that the system will stay in its ground state throughout the evolution.
Ideally, the energy of the system at the end of the  annealing will be the ground state energy of the problem Hamiltonian.

For this study, we use a 5000+ qubit quantum annealer called Advantage 4.1, accessed through a cloud-based service.
This quantum processing unit (QPU) features around 35,000 pair-wise couplers between qubits, arranged in a Pegasus topology of size 16 ($P_{16}$)~\cite{Dwave-QPU}.
In addition to specifying the QUBO problem, we can control the behavior of the QPU by setting the chain strength (see \Eq{rcs} below), annealing schedule, and the number of independent annealing runs per event.
The lowest energy found among the independent runs gives the final reported value.

A key step in programming a quantum annealer is finding an embedding of the QUBO problem onto the working graph of the QPU~\cite{choi2008minorembedding} (i.e.~the subset of the full chip graph excluding malfunctioning qubits and couplers).
This is a challenging problem in itself~\cite{choi2008minorembedding, Okada_2019}, often relying on heuristic methods and requiring several tries until a useful embedding is found.

\subsection{Qubit Chains}

The thrust QUBO problem in \Eq{qubo} requires a fully connected graph with couplers connecting every pair of qubits.
Since the $P_{16}$ topology does not allow for full connectivity, we must construct qubit chains.
A qubit chain is a group of physical qubits coupled together strongly enough to behave as a single logical qubit.
This technique increases the connectivity of the logical qubits, at the expense of having many fewer logical qubits than physical qubits.
For the D-Wave Advantage 4.1 QPU, the Pegasus topology can embed a fully connected graph with a maximum of 124 nodes.
A graph of this size requires chains of length 7.

To ensure that a qubit chain acts like single logical qubit, the strength of the coupling in the chain has to be determined empirically.
If the coupling is too weak, the chains will break, meaning that the logical qubits no longer behave as single units.
While one can apply post-processing techniques (e.g.\ majority vote) to determine the appropriate values for the logical qubits, results obtained in this way can have higher energy than the target ground state.
On the other hand, if the coupling strength is set too high, the chain couplers will dominate over the couplers in the interactions of the QUBO problem.
This produces a new logical problem that is more concerned with keeping chains consistent, and no longer yields a solution to the original QUBO problem.

A reasonable starting point for optimizing the chain strength is to set it to the largest coupling strength of the original QUBO problem~\cite{calaza2021garden}.
We define the \textit{relative chain strength} (RCS) as
\begin{equation}
    \text{RCS} = \frac{\text{ACS}}{\max_{ij}|Q_{ij}|},
    \label{eq:rcs}
\end{equation}
where $Q_{ij}$ is the QUBO matrix from \Eq{qubo_singlecone}, and ACS is the absolute chain strength that is used as an input parameter when calling the solver.
We set the default chain strength to RCS = 1 in \Sec{default_results}, and then explore alternative choices in \Sec{chain_results}.

Note that D-Wave's QPUs feature an $auto\_scale$ function, which rescales the coupling strength values in the problem to a range between $-1$ and $+1$.
This enables the user to submit problems with values outside these ranges and have the system automatically scale them to fit.
We perform this scaling separately for each thrust calculation.

\subsection{Sample Persistence Variable Reduction}
\label{sec:SPVAR_def}

When solving optimization problems, it is common to run a heuristic solver multiple times and keep the best solution found.
Suppose all of the found solutions are aggregated into an ensemble.
In that case, one could ask if there is any additional information to be gained by analyzing the ensemble as a whole, aside from the solution with the best value.

Previous studies have shown that it is indeed possible to use the ensemble more efficiently.
Consider fixing the variables that have the same value in a large number of the solutions obtained.
With these variables fixed, the remaining problem tends to be much smaller and simpler to solve.
This \textit{sample persistence variable reduction} (SPVAR) algorithm \cite{Karimi2017} has been reported to significantly increase the success rate in finding the best-known energy in quantum annealing applications.

In \Sec{SPVAR_results}, we consider a modified version of the SPVAR algorithm: the \textit{multi-start SPVAR} \cite{Karimi2017_2} shown in Algorithm~\ref{algorithm:multi_start_SPVAR}.
This approach iteratively fixes the value of a large portion of the variables to values that have a high probability of being optimal.

\begin{algorithm}[H]  
\footnotesize
\begin{algorithmic}
\Require QUBO problem, \textit{num\_reads}, 
\State \textit{fixing\_threshold}, \textit{elite\_threshold}, \textit{num\_starts} \\
\For{each start of \textit{num\textunderscore starts}}
        \State Obtain sample of \textit{num\textunderscore reads}
        \State Record energies from sample
        \State Narrow down solutions to \textit{elite\textunderscore threshold} percentile
        \State Find mean value of each variable in all solutions
        \State Fix variables for which mean absolute value is larger than \textit{fixing\textunderscore threshold}
\EndFor \\
\Return Recorded energies, and a mapping from fixed variables to values to which they were fixed
\end{algorithmic}
\caption{Multi-start SPVAR~\cite{Karimi2017_2}} 
\label{algorithm:multi_start_SPVAR}
\end{algorithm}

\subsection{Reverse Annealing}

By default, D-Wave's QPUs are initialized such that each of the $N$ qubits are in a uniform superposition of 0 and 1. 
With the usual forward schedule of quantum annealing, one starts with a high traverse field, which gradually decreases through the annealing process.
In this way, the quantum annealer performs a ``global'' search at the beginning of the annealing schedule when the transverse field is strong.
As the transverse field gets weaker, the search gets more ``local''.

D-Wave recently introduced a \textit{reverse annealing} feature~\cite{Dwave-RA} that allows the system to be initialized with a known classical solution.
The motivation for reverse annealing is to better explore the local space around a candidate solution to potentially find a state with even lower energy.
In reverse annealing, the system starts in a classical configuration defined by the user (e.g.~the result after standard forward annealing), then the transverse field is gradually increased (hence the name reverse annealing), after which the transverse field is gradually decreased again.
For some problems, it has been shown that optimizing the annealing schedule in this way results in better performance, by balancing global exploration of the state space with explorations in the vicinity of local optima.
Compared to the default forward approach, we indeed find somewhat better performance with reverse annealing in \Sec{reverse_results}.

\subsection{Seeded-Axis Iterative Search}
\label{sec:seed}

As discussed in \Sec{iterative}, it is possible to iteratively approximate thrust starting from an initial seed axis.
This classical algorithm is significantly faster than solving the exact optimization problem, though it only guarantees finding a local minimum.
It typically converges in a handful of iterations even for a large number of particles, much smaller than the $O(N^2)$ possible hemisphere partitions.

In \Sec{seed_results}, we apply this iterative approach to two choices of seed axis.
The first is using the solution obtained from the D-Wave Advantage QPU with default settings.
The second is using the sphericity axis found from \Eq{sph} with $r=1$.
While the iterative approach only finds a local minimum of thrust, it often does better than the plain annealing result with minimal additional computational overhead.

\subsection{D-Wave Classical Solvers}

Finally, in addition to hardware QPUs like Advantage (and its predecessor DW2000Q), D-Wave offers access to classical QUBO solvers.
Current QPUs have a limited number of qubits which might not be sufficient to solve problems at a real-world scale, but classical solvers can overcome this limitation.
We can also use classical solvers to benchmark the quantum annealing performance.
In \Sec{classical_results}, we present results obtained using the \textit{simulated annealing} algorithm \cite{simann}, as implemented in D-Wave's Ocean Software Development Kit version 3.4.1 with the default settings.

\begin{figure*}[t]
    \centering
    \includegraphics[width=\linewidth]{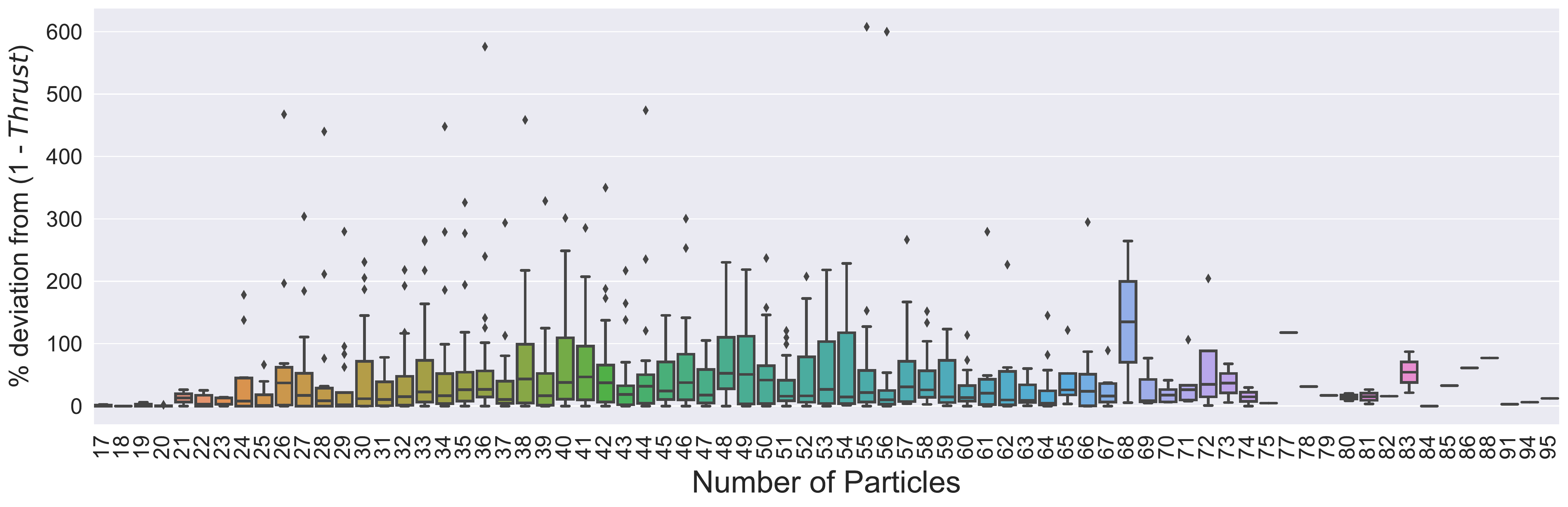}\\
    \caption{
    Quantum annealing results using the default parameters of the D-Wave Advantage 4.1 QPU.
    Shown is the percent deviation from the target value of one-minus-thrust, as a function of the number of particles.
    The box plots represent the median as a solid black line for each bin, as well as the first and third quartiles.
    Outlier points are displayed as black diamonds.}
    \label{fig:dwavedef}
\end{figure*}

\section{Results}
\label{sec:results}

We now present thrust results obtained from quantum annealing on the D-Wave Advantage 4.1 QPU and compare to hybrid and classical algorithms.
The dataset used for this benchmarking study consists of 1000 events from the 
process:
\begin{equation}
    e^{+}e^{-}\rightarrow\gamma^{*}/Z^{0}\rightarrow q\bar{q},
\end{equation}
generated in \textsc{Pythia} 8.303 with the default settings.
Events were generated at a center-of-mass energy of 91.1876 GeV (i.e.\ on the $Z$ pole).
We do not include the effects of detector acceptance or smearing in this study.

As discussed at the end of \Sec{thrust_definition}, we use the built-in thrust algorithm in \textsc{Pythia} to compute the exact thrust partition and value.

The generated events feature a varying number of particles, which allows us to study the performance of the Advantage system as a function of problem size.
In the 1000 events generated, the number of particles ranges from 17 to 95.
In our study, we found that the number of physical qubits needed to embed the QUBO problem scales as the number of particles on the event to roughly the power of 1.82.
We see that even though the Advantage system has many more physical qubits than the average number of particles in an electron-positron collision, the requirement in \Eq{qubo} of a fully connected graph limits the size of the problem we can solve.

\subsection{Default D-Wave Results}
\label{sec:default_results}

\begin{figure*}[p]
\centering
\subfloat[]{
\includegraphics[width=.75\columnwidth]{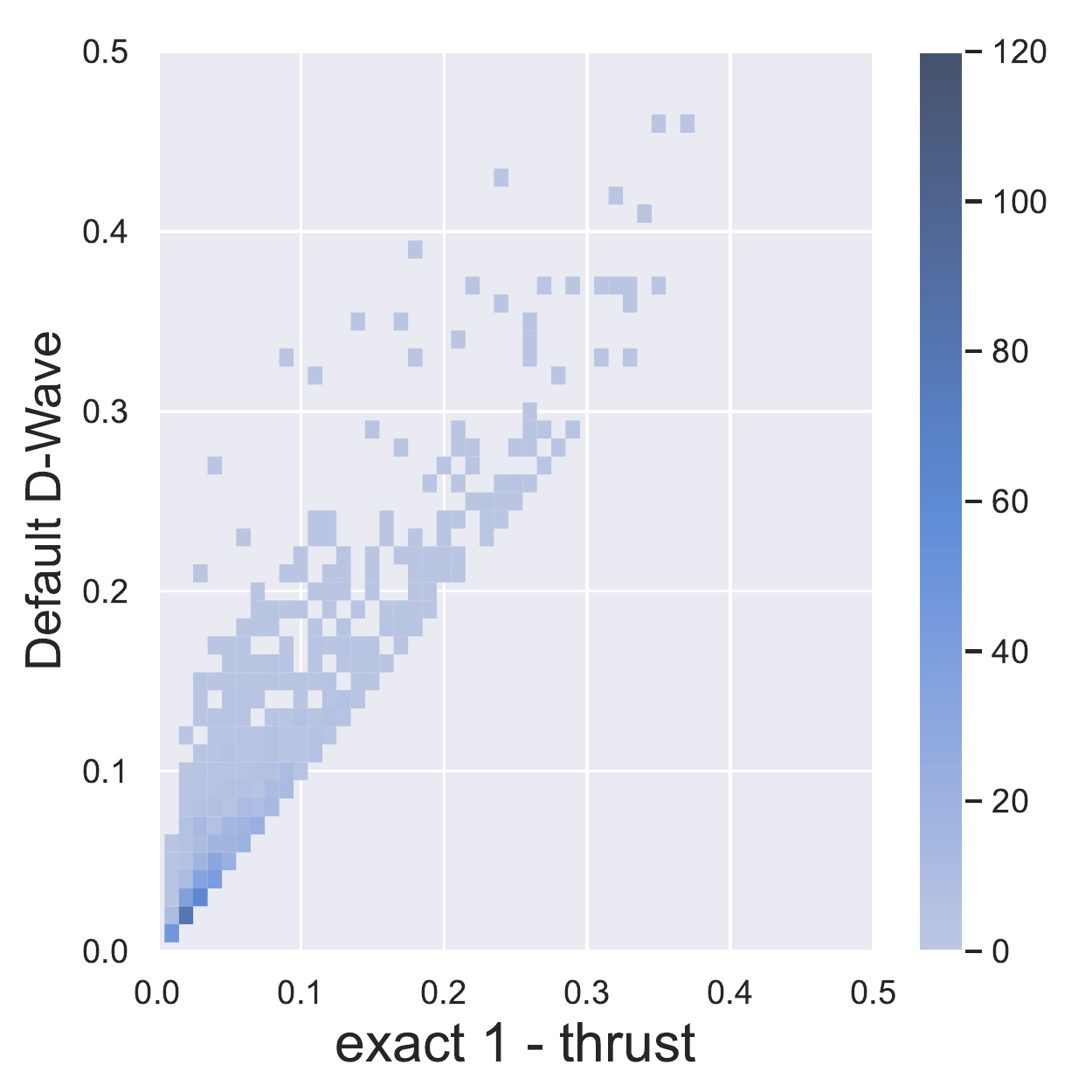}
\label{fig:dwaveresults_default}
}
\subfloat[]{
\includegraphics[width=.75\columnwidth]{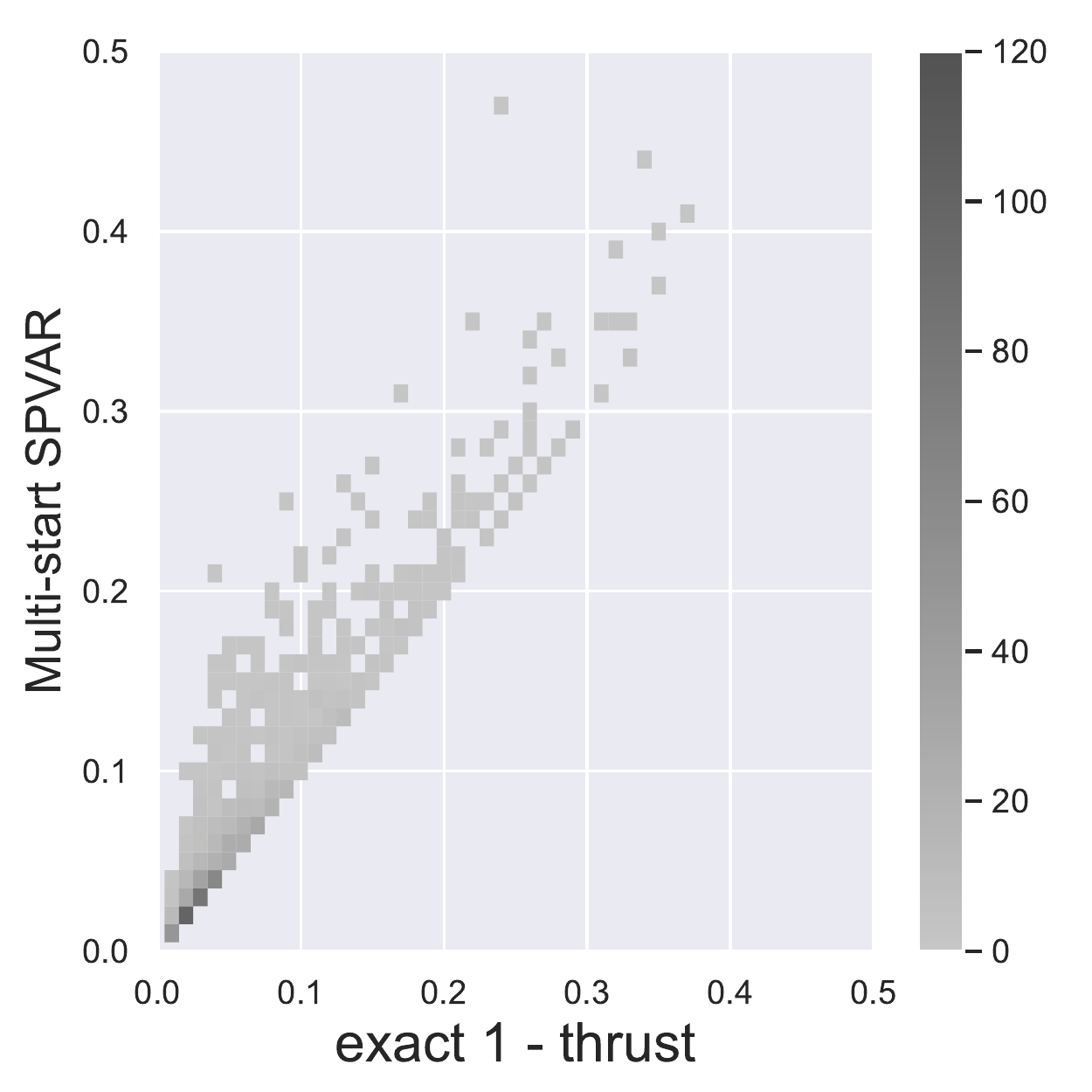}
\label{fig:iter_rem}
}\\
\subfloat[]{
\includegraphics[width=.75\columnwidth]{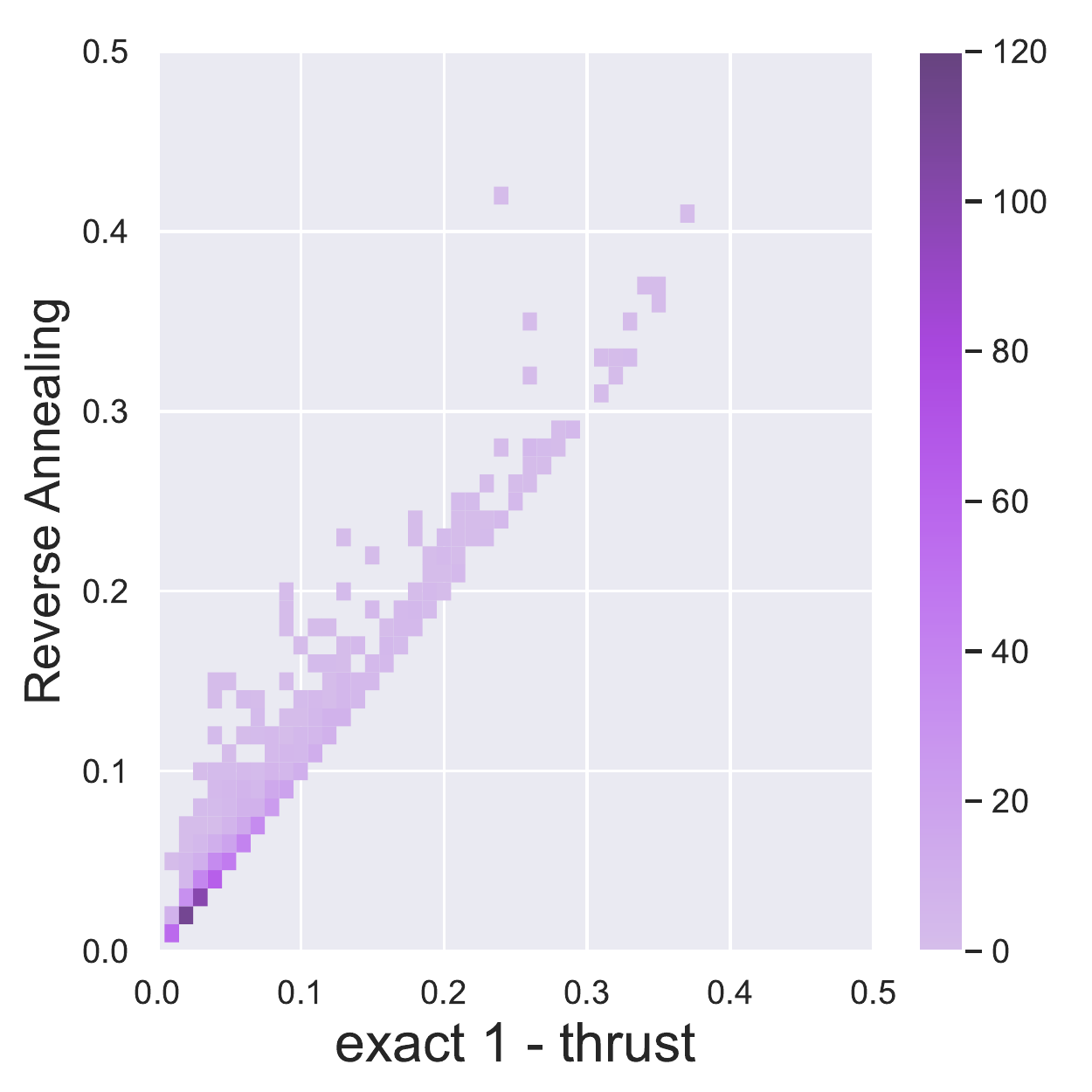}
\label{fig:revan}
}
\subfloat[]{
\includegraphics[width=.75\columnwidth]{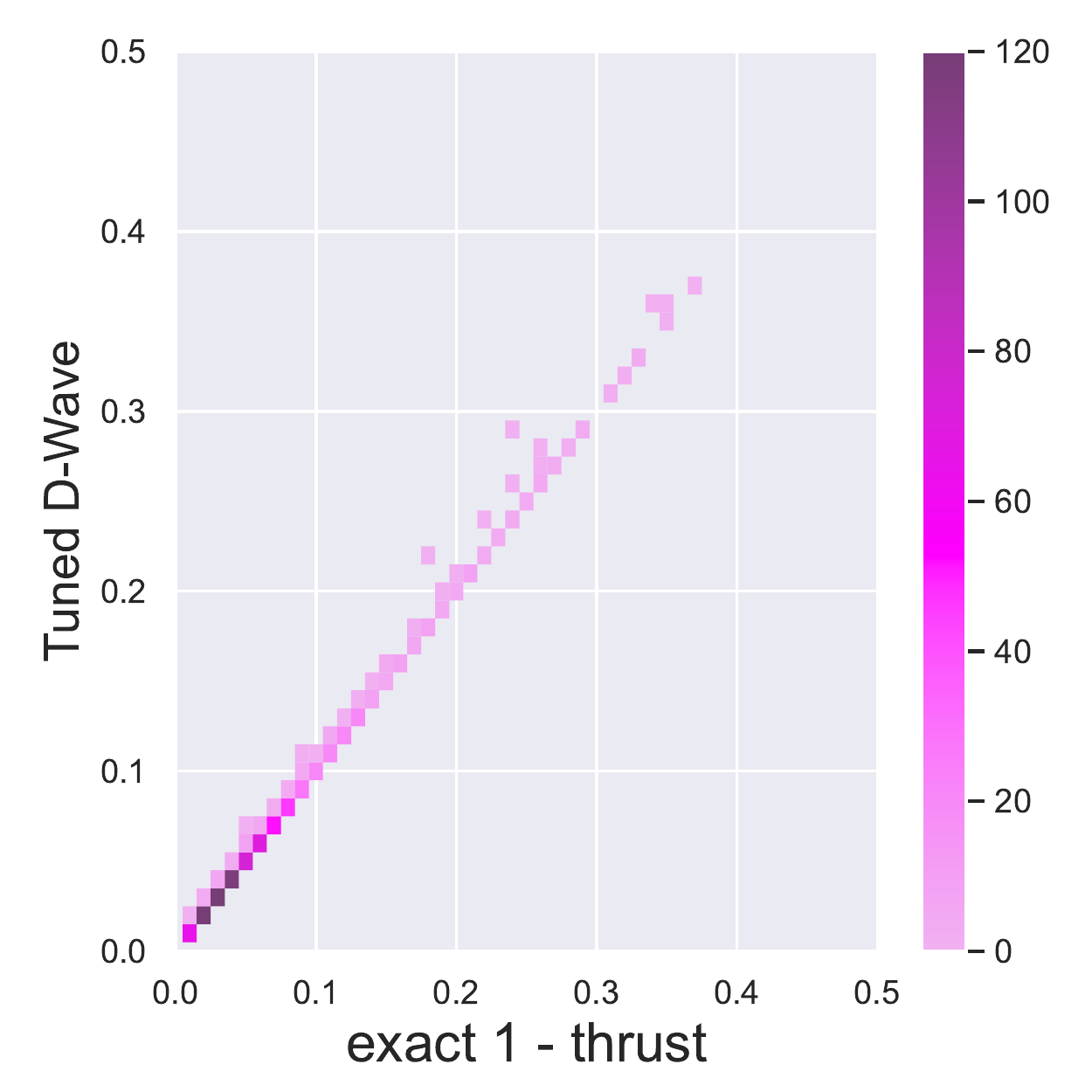}
\label{fig:dwaveresults_tuned}
}\\
\subfloat[]{
\includegraphics[width=.75\columnwidth]{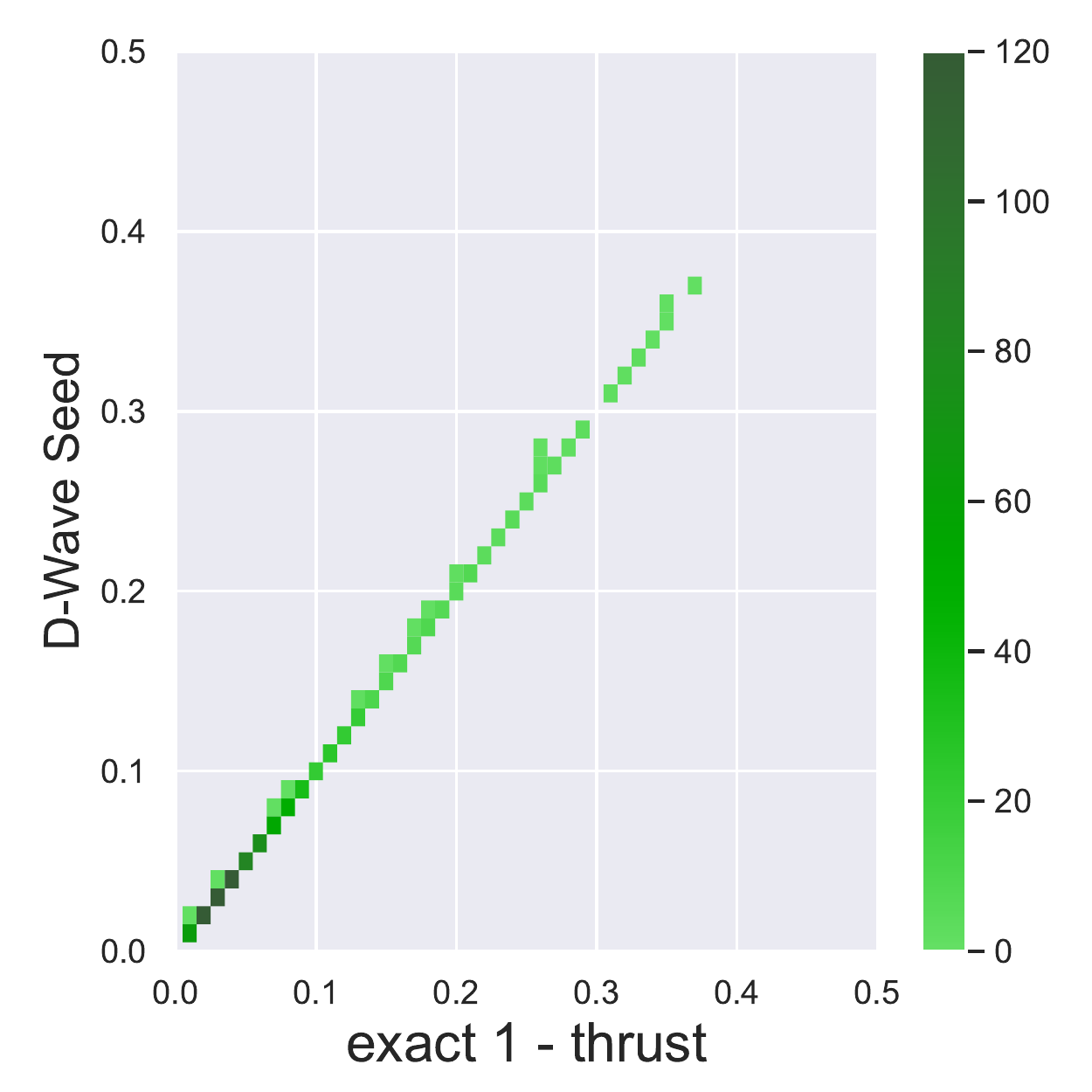}
\label{fig:seedresults}
}
\subfloat[]{
\includegraphics[width=.75\columnwidth]{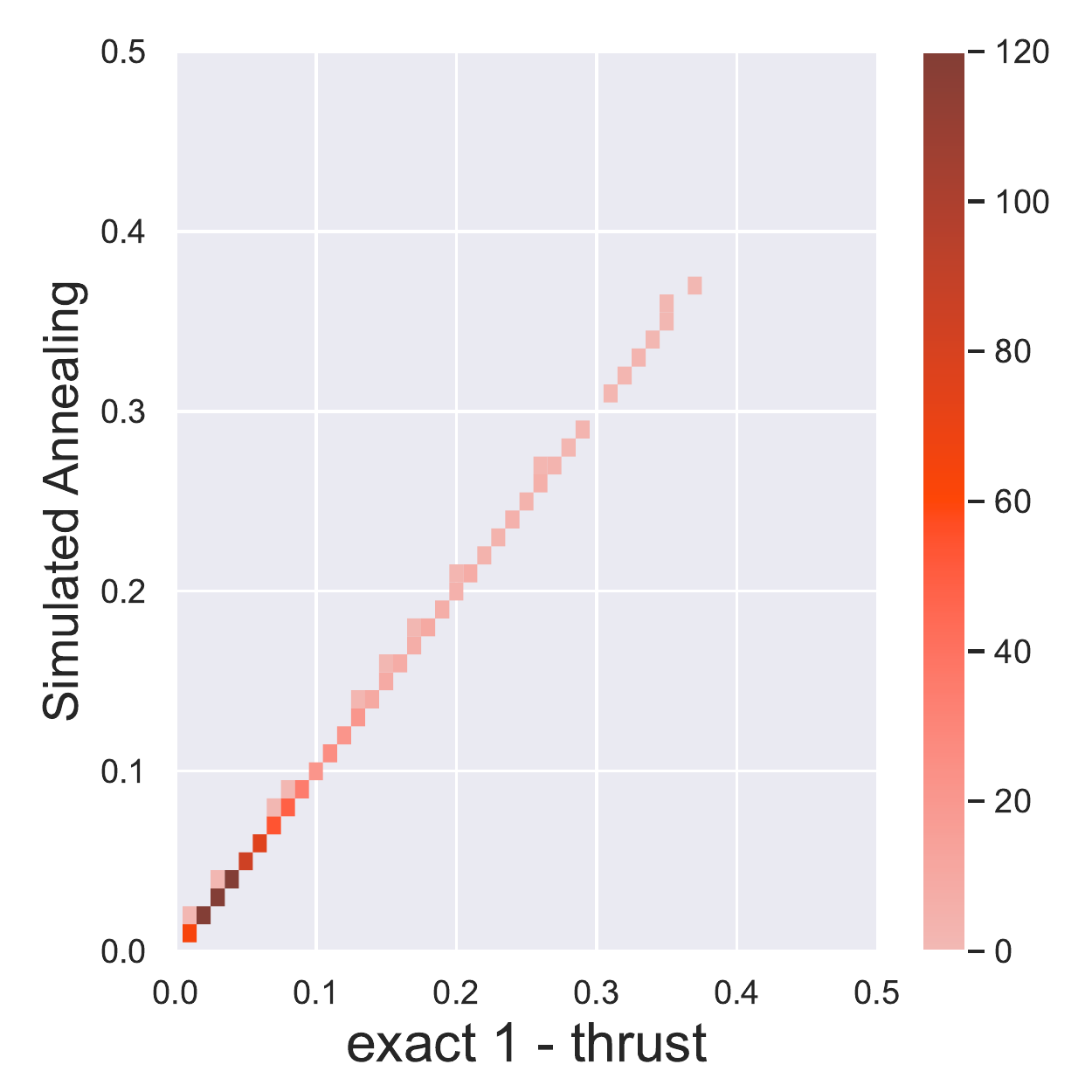}
\label{fig:saresults}
}
\caption{
Correlation between the one-minus-thrust value found using heuristic methods (vertical axis) and the exact method (horizontal axis).
Results from the Advantage QPU with default settings are shown for (a) forward annealing, (b) the SPVAR algorithm, and (c) reverse annealing.
Additional results are for (d) the Advantage QPU with tuned setting, (e) using the forward annealing results without tuning as a seed for classical iterative improvement, and (f) classical simulated annealing.
}
\end{figure*}

The default annealing parameters for the D-Wave Advantage 4.1 QPU are:
\begin{itemize}
    \item Relative chain strength (RCS):  1.0; 
    \item Annealing time: 20 $\mu$s;
    \item Number of runs per event (\textit{num\_reads}):  100.
\end{itemize}
With these parameters, one event would minimally take 2 milliseconds to process.
Including the overhead of embedding the QUBO problem on the Advantage system, the runtime is on the order of 0.1 to 100 seconds depending on the number of particles.
The reported solution is the best one found from the 100 annealing runs.

As shown in \Fig{dwavedef}, these default parameters yield relatively poor performance, with upwards of 50\% deviations from the target value of one-minus-thrust ($1-T$).
The performance is relatively independent of the number of particles.
Note that the slight apparent gain in performance with more particles is an artifact, since more particles correspond to larger values of $1-T$ and therefore smaller percentage deviations.
In \Fig{dwaveresults_default}, we plot the correlation between the found thrust value and the exact thrust value.
While there is a cluster of events along the diagonal with good behavior, there are large tails where even after 100 runs, the QPU does not find the correct solution.

\subsection{Sample Persistence Variable Reduction}
\label{sec:SPVAR_results}

To try to improve the performance, we test the multi-start SPVAR algorithm described in \Sec{SPVAR_def}.
We use the default QPU parameters from \Sec{default_results} and set the SPVAR parameters as:
\begin{itemize}
    \item Number of iterations (\textit{num\_starts}):  10; 
    \item Solutions kept for analysis (\textit{elite\_threshold}): 100\% but poor solutions removed as described below;
    \item Threshold for fixing qubit (\textit{fixing\_threshold}):  $0.65$.
\end{itemize}
Because of the multiple iterations, one event now minimally takes 20 milliseconds to process, though this is dwarfed by the computational overhead.

The SPVAR algorithm takes the set of non-optimal solutions from a given iteration and attempts to increase the success rate of successive iterations by fixing the value of variables to either $0$ or $1$.
For a sample size of 100 and a fixing threshold of $0.65$, if a variable is found to have the same value in more than 65 samples, the variable is fixed to that value.
Furthermore, after autoscaling the couplings, the energy associated to a given solution must be greater than the sum of the momenta of all particles in the event, divided by 6, for the sample to be considered in the count.
In \Fig{iter_rem}, we show the correlation between the SPVAR thrust values obtained after 10 iterations and the exact thrust values.
There is a modest improvement in the solutions obtained compared to those obtained through the default annealing, though at a higher computational cost.

\subsection{Reverse Annealing}
\label{sec:reverse_results}

We next test a modified annealing schedule, based on D-Wave's \textit{Reverse Advance Composite}.
This module allows the user to {reverse anneal} an initial sample through a sequence of annealing schedules.

An annealing schedule is specified as a list of $[t,s]$ pairs, in which time $t$ is given in microseconds from the run start and the normalized persistent current $s$ is given in the range $[0,1]$.
In this format, the default forward annealing corresponds to: 
\begin{equation}
\{[0.0, 0.0], [20.0, 1.0]\},
\end{equation}
which yields a linear ramp up of the persistent current.
For reverse annealing, we use the schedule:
\begin{equation}
\label{eq:reverseannealschedule}
\{[0.0, 1.0], [t, 0.5], [20.0, 1.0]\},
\end{equation}
for $t$ in $\{ 5, 10,15 \}$.
While the total annealing time per run is still 20 microseconds, we start from full current, reverse anneal to half current, and then forward anneal back up to full current.

For the first run in a submission, we start from a random solution.
Then, each subsequent run uses the best solution found thus far as its initial state.
We run 100 reverse anneals for each of the three schedules in \Eq{reverseannealschedule},  and show results from the best solution in \Fig{revan}.
We find a modest improvement in the quality of the solutions  compared to the default method, for essentially the same computational cost.

\subsection{Tuning the D-Wave QPU}

Given that the SPVAR and reverse annealing techniques did not lead to a substantial improvement in the quality of the annealing results, we now turn to trying to tune the annealing parameters for the problem we are trying to solve.
We explore the impact of changing the chain strength, annealing time, and sample size.
While we do not attempt an exhaustive optimization, we achieve significantly better results using tuned parameters, as shown in \Fig{dwaveresults_tuned}.

\subsubsection{Chain Strength Scan}
\label{sec:chain_results}

\begin{figure}[t]
    \centering
    \includegraphics[width=.98\linewidth]{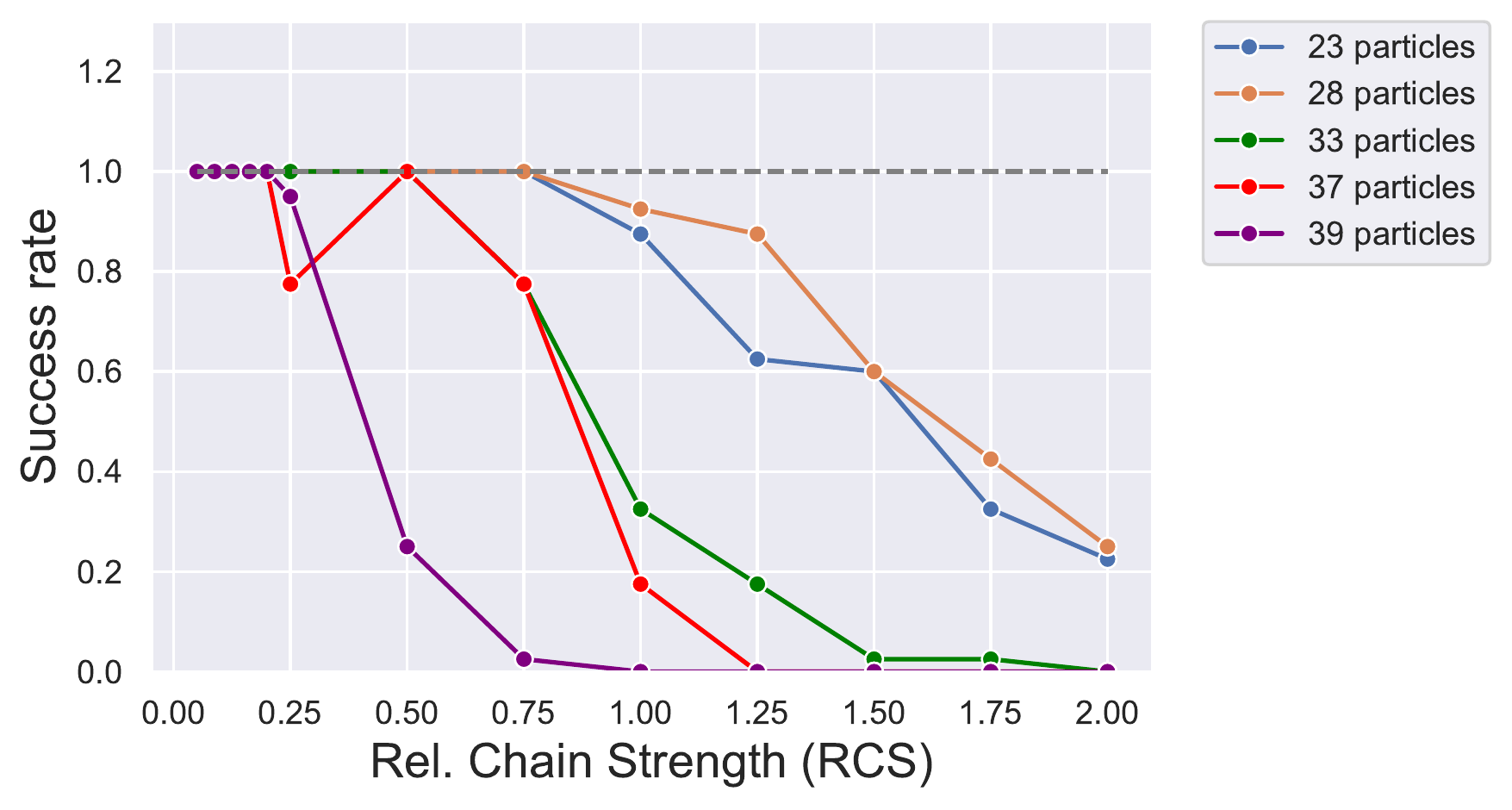}
    \caption{
    Success rate as a function of the RCS in \Eq{rcs} for five different events with increasing number of particles:
    23 (blue), 28 (orange), 33 (green),  37 (red), and 39 (purple).
    For each event, we report the success rate among 40 executions, where each execution consists of $10^{4}$ annealing runs with the default annealing time of 20 $\mu$s.
    }
    \label{fig:ch_str_Scan}
\end{figure}

The RCS in \Eq{rcs} controls the degree to which chained physical qubits behave as one logical qubit.
Larger values of RCS preserve logical qubit coherence at the expense of making it harder to minimize the QUBO loss.
Smaller values of RCS allow qubit chains to break, leading to inconsistent results.

To optimize the choice of RCS, we consider five representative events with 23, 28, 33, 37, and 39 particles, respectively.
For each event, we consider 13 different values of the RCS, ranging from 0.05 to 0.20, increasing in steps of 0.0375, and from 0.25 to 2.00, increasing in steps of 0.25.
To emphasize the impact of RCS on finding solutions, we increase the number of runs per event to $10^4$, compared to the default of 100.

The results of the RCS scan are shown in \Fig{ch_str_Scan}, where the success rate is determined by performing 40 independent executions and counting the fraction of executions where the solver returns the exact result.
We see that we can improve the success rate by tuning the RCS value for a given problem.
The overall trend is that lower RCS values yield higher success rates, implying that chain coherence is less important than chain flexibility.
For larger problems, there is a smaller range of RCS values for which the success rate is close to unity.
We tested an event with 45 particles that had a success rate of 0 regardless of the RCS value.
The success rate for most events is largely unchanged for RCS values at or below 0.2, so we take RCS = 0.2 for our tuned parameter value, compared to the default of RCS = 1.0.

\subsubsection{Annealing Time Scan}

\begin{figure}[t]
\centering
\includegraphics[width=.95\linewidth]{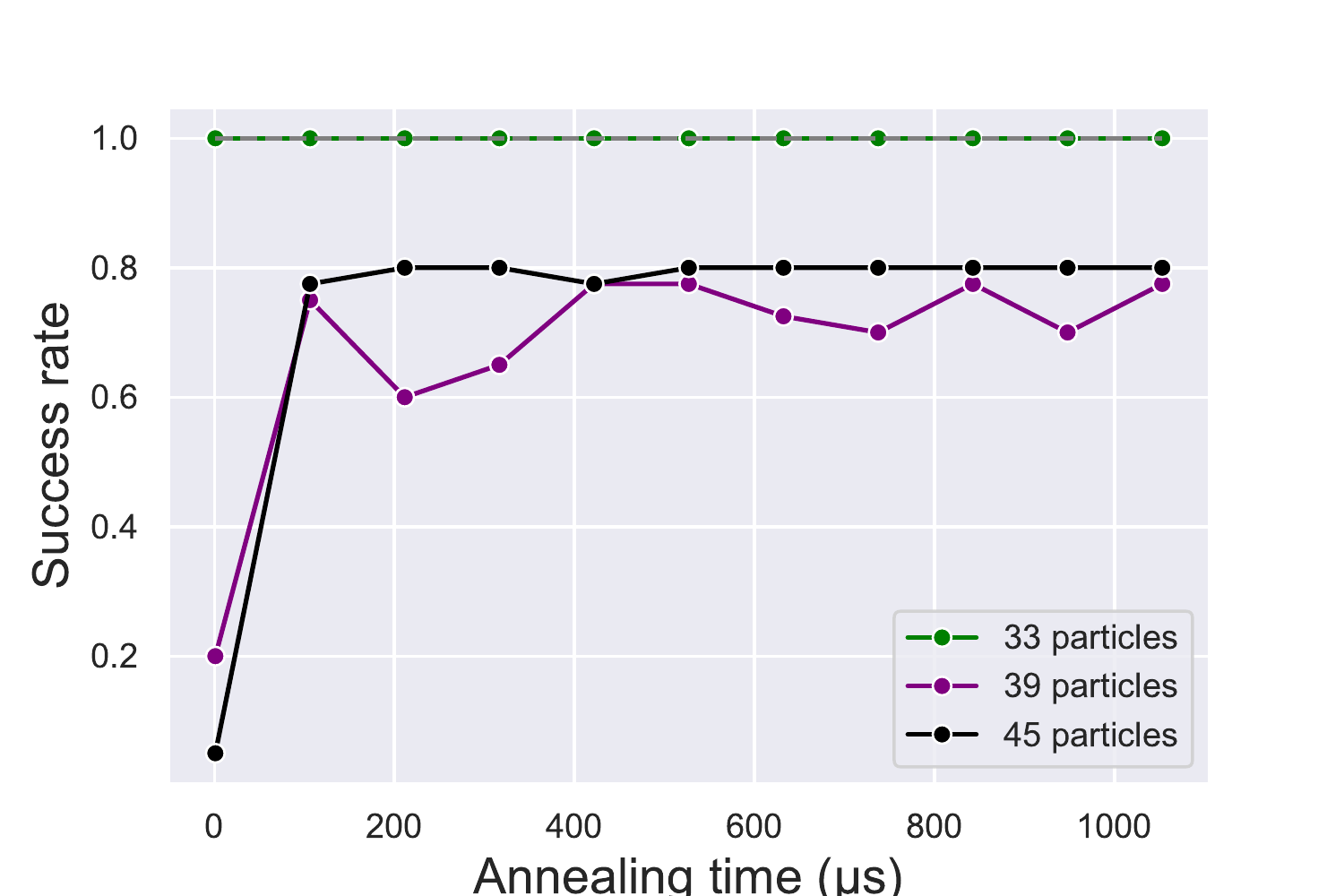}\\
\caption{
Success rate as a function of annealing time for three events with increasing number of particles: 33 (green), 39 (purple), and 45 (black).
For each event, we report the success rate among 40 executions, where each execution consists of $10^{4}$ annealing runs with RCS = 0.2
}
\label{fig:anneal_scan}
\end{figure}

In order to increase the success rate for large problem sizes, we need to tune another crucial parameter: the annealing time.
Longer annealing times increase the accuracy of the adiabatic approximation that underlies quantum annealing.

\begin{figure*}[t]
\centering
\includegraphics[width=.95\linewidth]{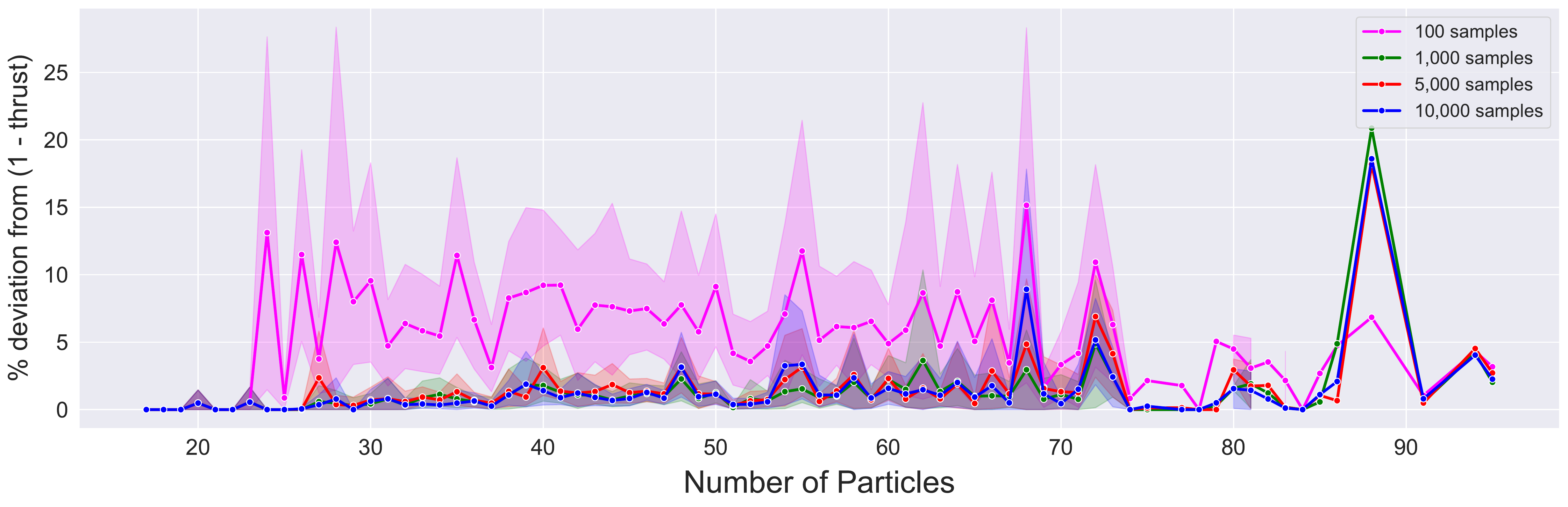}\\
\caption{
Average percentage deviation from $1 - T$ as a function of number of particles in the event, acting on the full dataset, for increasing number of annealing samples: 100 (fuchsia), 1,000 (green), 5,000 (red), and 10,000 (blue).
For each problem, the RCS was set to $0.2$ and the annealing time to 100 $\mu$s.}
\label{fig:numreads_scan}
\end{figure*}

We selected three events from the chain strength scan study, containing 33, 39, and 45 particles, respectively.
For each event, we perform an annealing time scan using ten values evenly spaced linearly in the range from 1 to 1000 $\mu$s.
(The maximum annealing time for the Advantage QPU system is 2000 $\mu$s).
We set RCS = 0.2, which, according to \Fig{ch_str_Scan}, is optimal for the 39 particle event, allowing us to emphasize the impact of annealing time.
As with the chain strength scan, we set the number of runs per event to $10^{4}$.

As shown in \Fig{anneal_scan}, even with an annealing time of 1 $\mu$s, the QPU already saturates the success rate for the 33 particle problem.
For the 39 and 45 particle problems, the success rate rises to around 75\% for an annealing time 100$\mu$s, with no gain in performance for longer runs.
Of course, longer annealing times for a fixed number of runs result in longer execution times.
Thus, one needs to consider whether the increase in execution time resulting from setting a longer annealing time value would not be better spent in generating additional runs with shorter annealing times.
We decided not to attempt a more refined optimization of the annealing time, and we fix our tuned annealing time to 100 $\mu$s, compared to the default value of 20 $\mu$s.

\subsubsection{Sample Size Scan}

Our last scan is aimed at determining the right number of annealing runs needed to obtain reasonable performance.
Using the full dataset, we performed independent runs of sizes 100, 1k, 5k, and 10k.
The results are shown in \Fig{numreads_scan}, where we compare the average percent deviation from one-minus-thrust as a function of the number of particles.
While there is considerable improvement going from 100 runs to 1k runs, the performance is very similar for larger run sizes.
We therefore set the tuned sample size to 1k runs, compared to the default value of 100.

\subsubsection{Tuned Results}
\label{sec:tuned_results}

In summary, we choose our tuned annealing parameters to be:
\begin{itemize}
    \item Relative chain strength:  0.2;
    \item Annealing time: 100 $\mu$s;
    \item Number of runs per event:  1000.
\end{itemize}
Compared to the default in \Sec{default_results}, this represents a factor of around 50 increase in minimum runtime per event.
That said, the overhead of embedding the QUBO problem onto the Advantage system dominates the runtime.
The results are shown in \Fig{dwaveresults_tuned}, which represent a considerable improvement to those in \Fig{dwaveresults_default}.

\subsection{Seeded-Axis Iterative Search}
\label{sec:seed_results}

\begin{figure}
    \centering
    \includegraphics[width=.9\linewidth]{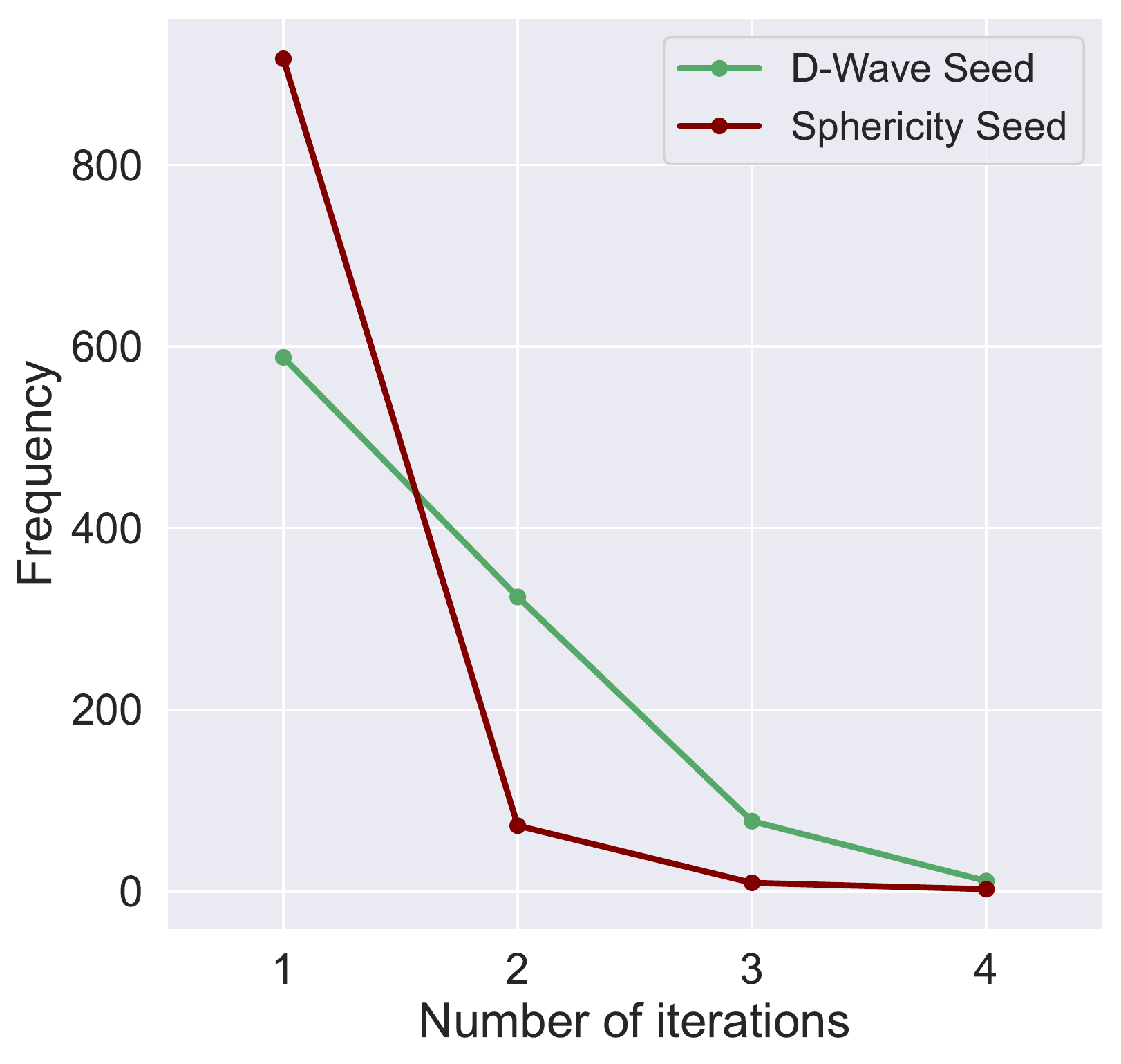}
    \caption{
    Number of iterations needed to reach a local optimum when using a seed from the default D-Wave annealing (red) compared to a seed from the linearized sphericity tensor (blue).}
    \label{fig:niter}
\end{figure}

Having tuned the Advantage QPU system, we now compare its performance to hybrid and classical solving strategies.
As described in \Sec{seed}, starting from seed axis, one can iteratively find an improved value of thrust.
We explore two choices of seeds:
\begin{itemize}
    \item \textit{Hybrid quantum/classical}:  the axis from the default D-Wave annealing in \Sec{default_results};
    \item \textit{Classical only}:  The linearized sphericity axis defined in \Sec{sphericity}.
\end{itemize}

In \Fig{seedresults}, we show the results of using the default D-Wave annealing as the seed, which is a hybrid quantum/classical approach.
Here, the seed axis is determined by taking the three-momentum sum of the particles clustered in QUBO formulation from \Eq{qubo}.
We find substantially better performance than the default D-wave results, with performance comparable to the tuned annealing.
Given the low computational overhead of the iterative approach, this suggests the value of hybrid strategies where quantum (reverse) annealing could be used in concert with classical iterative improvement.

In \Fig{niter}, we show the number of iterations needed to converge to a global minimum, comparing the D-Wave seed to the sphericity seed.
In general, the sphericity axis requires fewer iterations to converge to a local optimum.
In particular, for more than 90\% of the events, it takes at most one iteration to reach the true ground state starting from the sphericity seed. 
This suggests that the sphericity seed is a better starting point for the iterative approach, which we quantify further in \Sec{final_comparison}.

\subsection{D-Wave Classical Solver}
\label{sec:classical_results}

As a final comparison, we test the performance of classical simulated annealing.
As apparent from \Fig{saresults}, results obtained from this approach are very close to the true ground state solution obtained through the exact solver.
This serves as a sanity check on the formulation of the QUBO problem.
The runtime for classical simulating annealing is on the order of 10 ms to 100 ms per event, which is around 2 orders of magnitude faster than the quantum annealing approach when overhead is included.

\subsection{Performance Comparison}
\label{sec:final_comparison}

\begin{figure*}[p]
    \centering
    \includegraphics[width=.85\linewidth]{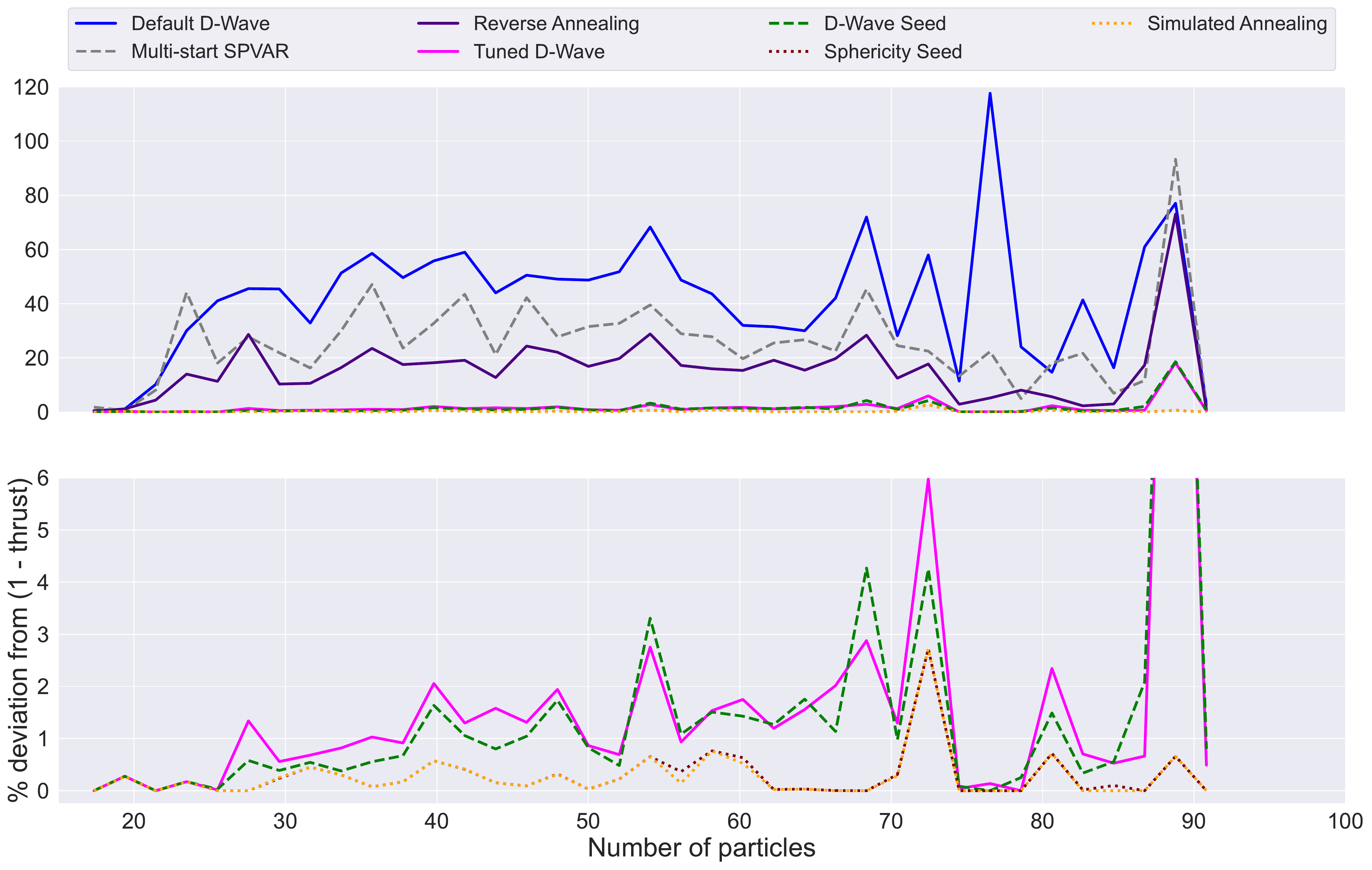}
    \caption{
    Average percentage deviation from the exact value of
    one-minus-thrust, as a function of the number of particles.
    Shown are the seven solving methods studied in this paper:  default D-Wave annealing (blue), multi-start SPVAR (grey), reverse annealing (purple), tuned D-Wave annealing (pink), iterative improvement from the default D-Wave seed (green), iterative improvement from the sphericity seed (red), and classical simulated annealing (orange).
    Quantum algorithms are displayed in solid lines, while hybrid and classical algorithms are displayed in dashed and dotted lines, respectively.
    The bottom plot is a vertical zoom in of the top plot to highlight the best performing algorithms.
    }
    \label{fig:npart}
\end{figure*}

\begin{figure*}[p]
    \centering
    \includegraphics[width=.85\linewidth]{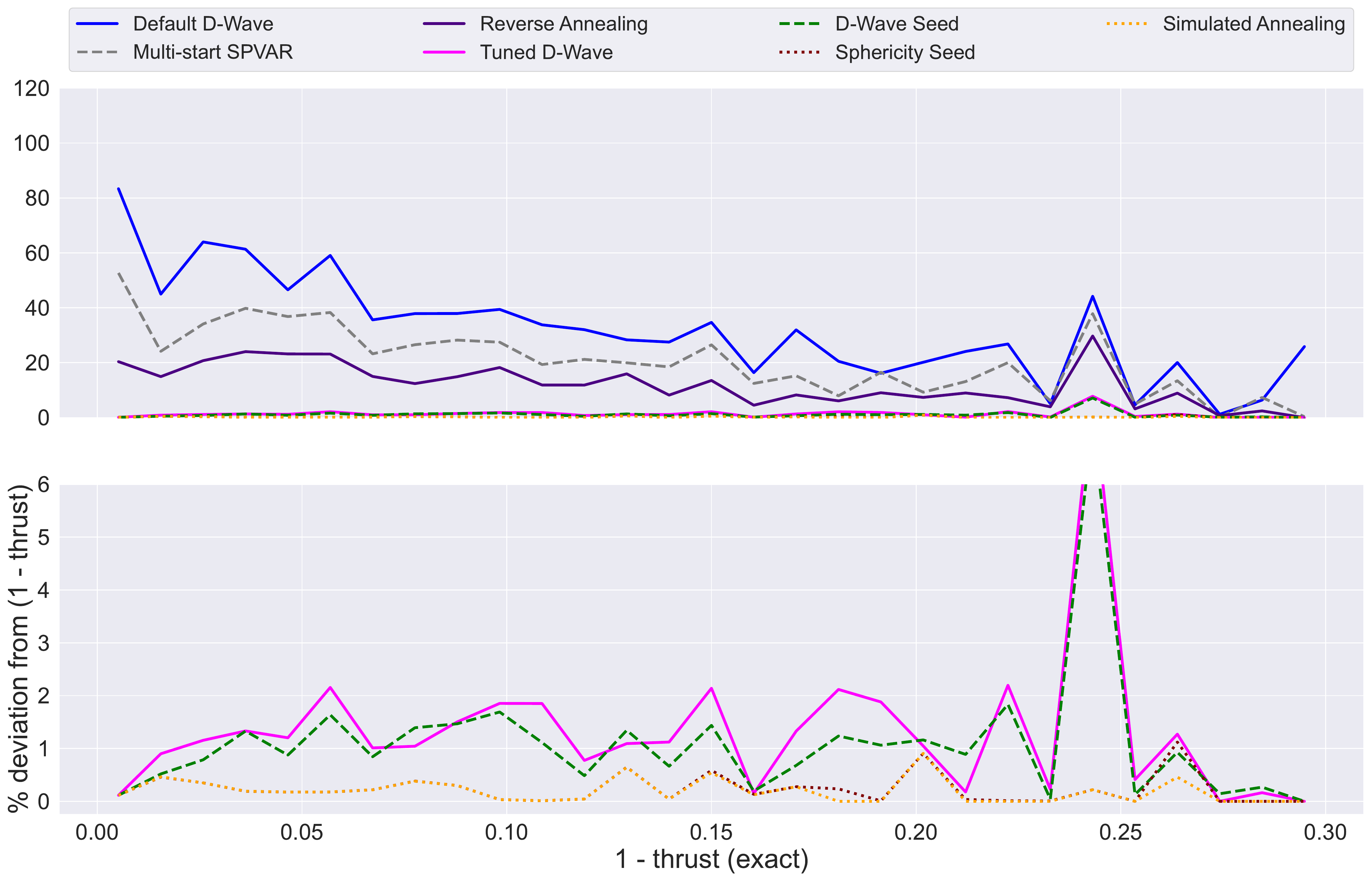}
    \caption{
    Same as \Fig{npart} but now plotted as a function of the exact one-minus-thrust.
    }
    \label{fig:one-minus-thrust-all}
\end{figure*}

We now summarize results for all of the approaches applied to the thrust QUBO problem, including quantum, hybrid, and classical techniques.
In \Fig{npart}, we plot the average percentage deviation in the found value of one-minus-thrust as a function of the number of particles in the event.
We show results as a function of one-minus-thrust in \Fig{one-minus-thrust-all}.

For the three untuned annealing strategies (default D-Wave, multi-start SPVAR, and reverse annealing), we see deviations ranging from 10\% to 50\%, depending on the precise algorithm used and target thrust value.
For fixed annealing time, algorithmic improvements like reverse annealing can yield substantially improved performance on this problem, but the gains are not sufficient to match the classical solvers.

The tuned annealing strategy exhibits a substantial improvement, with deviations now in the range of 1\% to 3\% percent, with a few outliers.
Intriguingly, comparable performance is exhibited for the hybrid quantum/classical approach of iterating from the untuned annealing seed.
This suggests that the tuned and untuned annealers often find the same basins of attraction, but the tuned annealer is able to get closer to the local optimum.

Thus, by tuning quantum annealers, or employing hybrid quantum/classical methods, we can roughly match the performance of classical heuristics for the thrust problem.
That said, the best performance is obtained from the two classical strategies: the iterative approach starting from the sphericity axis and simulated annealing.
For some (but not all) particle configurations, the thrust problem exhibits local minima, which the classical methods are able to more often avoid.
Nevertheless, the quantum and hybrid strategies find sensible thrust values, and there may be ways to improve the performance with further algorithmic refinements.

\section{Conclusion}
\label{sec:conclusion}

In this paper, we benchmarked thrust-based quantum annealing algorithms for jet clustering on the D-Wave Advantage 4.1 QPU.
Algorithmic improvements like multi-start SPVAR and reverse annealing showed some promise on this problem, but the biggest gains came from tuning the annealing parameters.
As expected, longer annealing times and more runs yielded improved performance.
Less obvious were the gains from reducing the chain strength, which suggests that further improvements could be obtained through more dynamic chain strength specifications.

Thrust is an interesting problem for benchmarking quantum algorithms for high-energy physics since so much is already known about its behavior.
The iterative classical heuristic is a well-known approach for estimating thrust, and we compared the performance starting from a quantum-derived seed versus a classical sphericity seed.
The fact that the classical seed performs better than the quantum seed suggests that quantum annealing sometimes gets trapped in local optima.
More investigations into hybrid quantum/classical methods could yield better approaches to addressing this issue.

The primary limitation of this study is that we had to restrict our attention to problems that could fit on the D-Wave QPU.
We did not observe any obvious trends with the number of particles, so we do not know how these results might extrapolate to future larger systems.
Given that smaller chain strength parameters were preferred, it would be interesting to study whether purposefully broken chains might yield comparable performance.
This would allow the embedding of larger problems at the expense of not having a consistent specification of a fully connected graph.

Overall, these results suggest that quantum annealing and hybrid quantum/classical algorithms are viable approaches to performing clustering tasks in high-energy physics.
We look forward to extensions of this work to multi-jet algorithms, with the hope of finding efficient ways to use quantum annealing to identifying jets in proton-proton events in conditions similar to those at the LHC.

\begin{acknowledgments}

We thank Jean-Roch Vlimant for helpful comments on a previous version of this manuscript.
We thank the University of Southern California for a grant of time on the D-Wave system at the USC-Lockheed Martin Quantum Computing Center at the Viterbi School of Engineering.
A.D. is supported by the Quantum Information Science Enabled Discovery (QuantISED) for High Energy Physics program at Oak Ridge National Laboratory under FWP ERKAP61, and by the Laboratory Directed Research and Development Program of Oak Ridge National Laboratory, managed by UT-Battelle, LLC, for the U.S.\ Department of Energy.
J.T. is supported by the U.S.\ Department of Energy (DOE) Office of High Energy Physics under contract DE-SC0012567, and by the DOE QuantISED program through the theory consortium ``Intersections of QIS and Theoretical Particle Physics'' at Fermilab (FNAL 20-17).

\end{acknowledgments}

\bibliography{quantum_thrust}

\appendix
\section{Additional Plots}

For completeness, in \Fig{per_devitation} we show the percent deviation from the target value of one-minus-thrust for six optimization schemes studied in this paper.
Note that the vertical axis differs substantially between each of these plots.

\begin{figure*}[p]
\centering
\subfloat[]{
\includegraphics[width=.95\linewidth]{Figures/percentage_DWaveExact.pdf}
\label{fig:per_dwaveresults_default}
}\\
\subfloat[]{
\includegraphics[width=.95\linewidth]{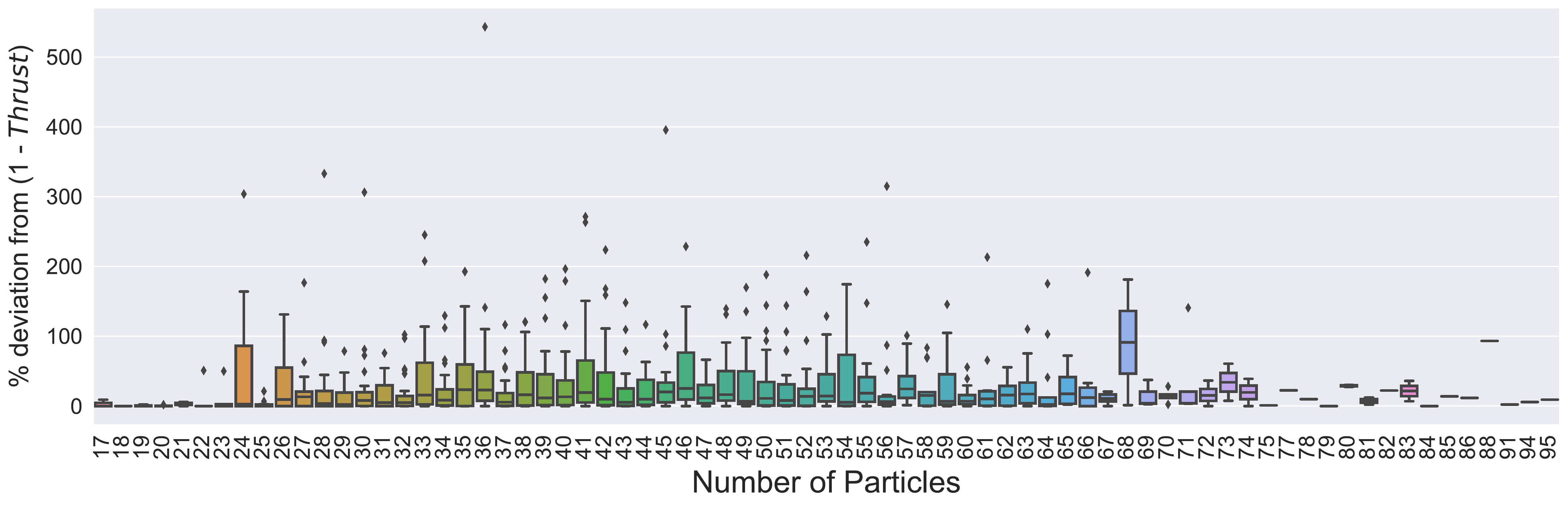}
\label{fig:per_iter_rem}
}\\
\subfloat[]{
\includegraphics[width=.95\linewidth]{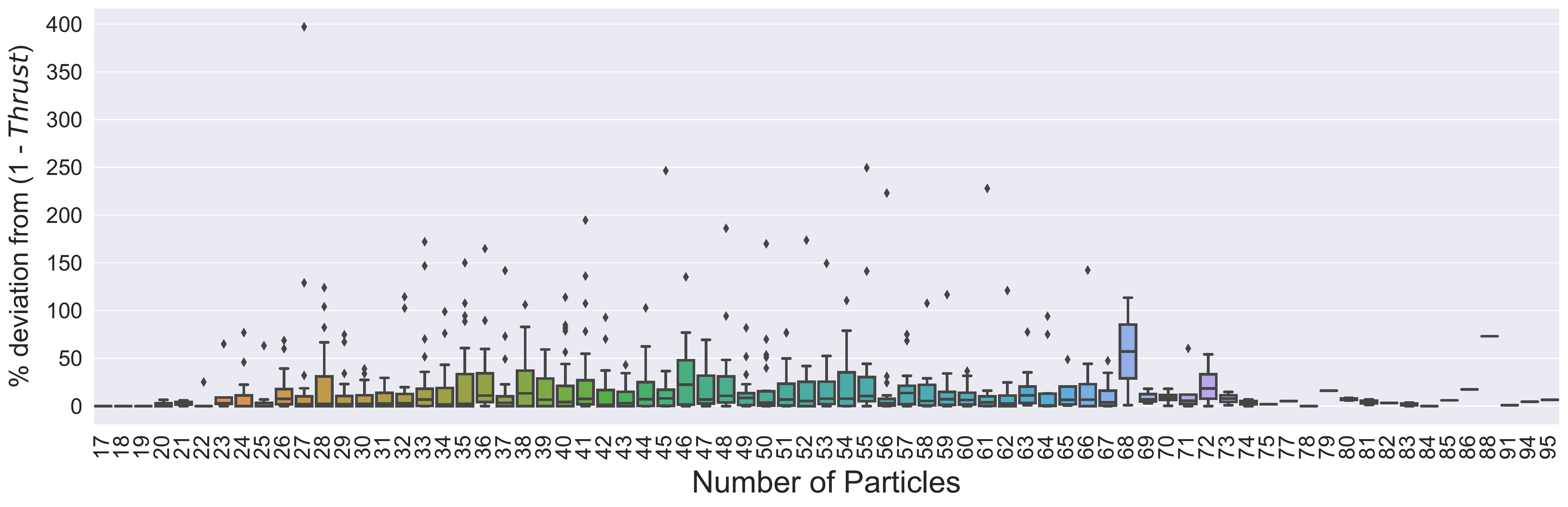}
\label{fig:per_revan}
}\\
\caption{Percent deviation from the target value of one-minus-thrust, as a function of number of particles. 
The box plots represent the median as a solid black line for each bin, as well as the first and third quartiles. Outlier points are displayed as black diamonds.
Results from the Advantage QPU with default settings are shown for (a) forward annealing (repeated from \Fig{dwavedef} for convenience), (b) the SPVAR algorithm, and (c) reverse annealing.
Continued on next page.
}
\label{fig:per_devitation}
\end{figure*}

\begin{figure*}[p]
\ContinuedFloat
\centering
\subfloat[]{
\includegraphics[width=.95\linewidth]{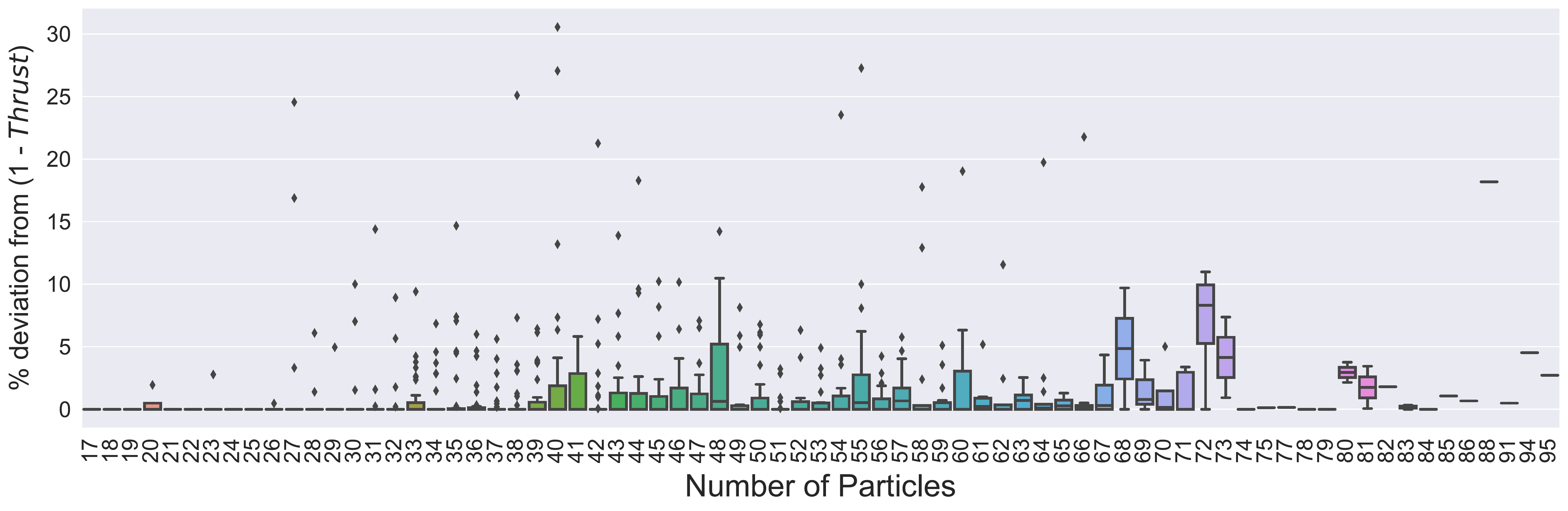}
\label{fig:per_dwaveresults_tuned}
}\\
\subfloat[]{

\includegraphics[width=.95\linewidth]{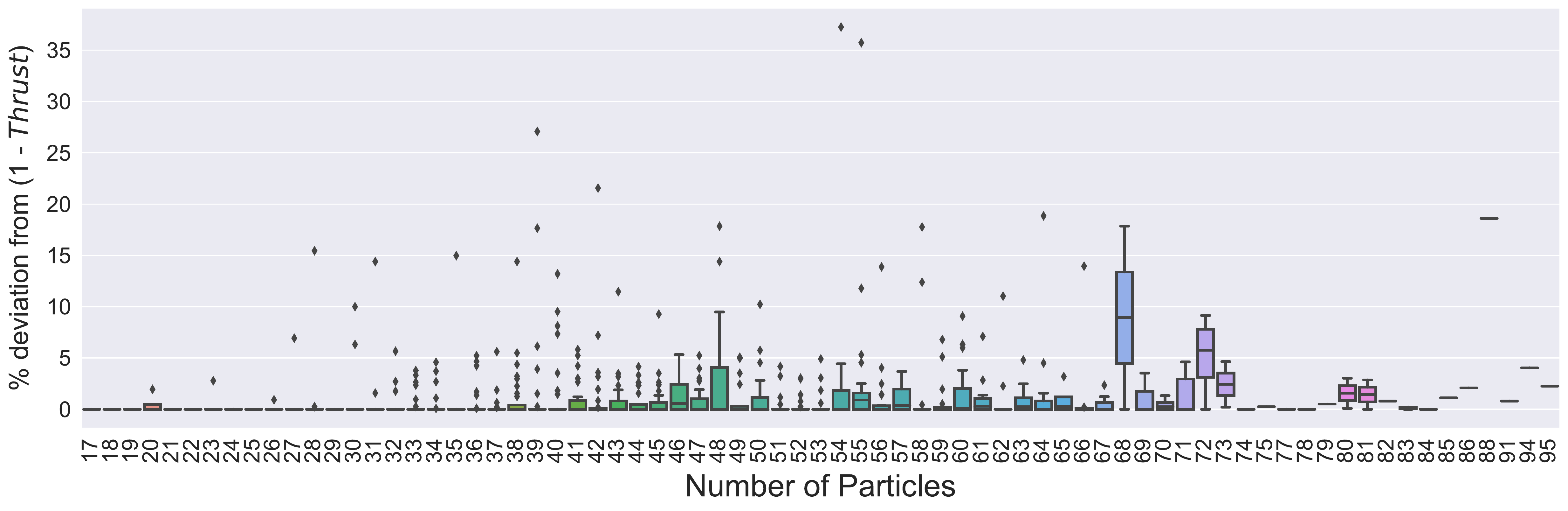}
\label{fig:per_seedresults}
}\\
\subfloat[]{
\includegraphics[width=.95\linewidth]{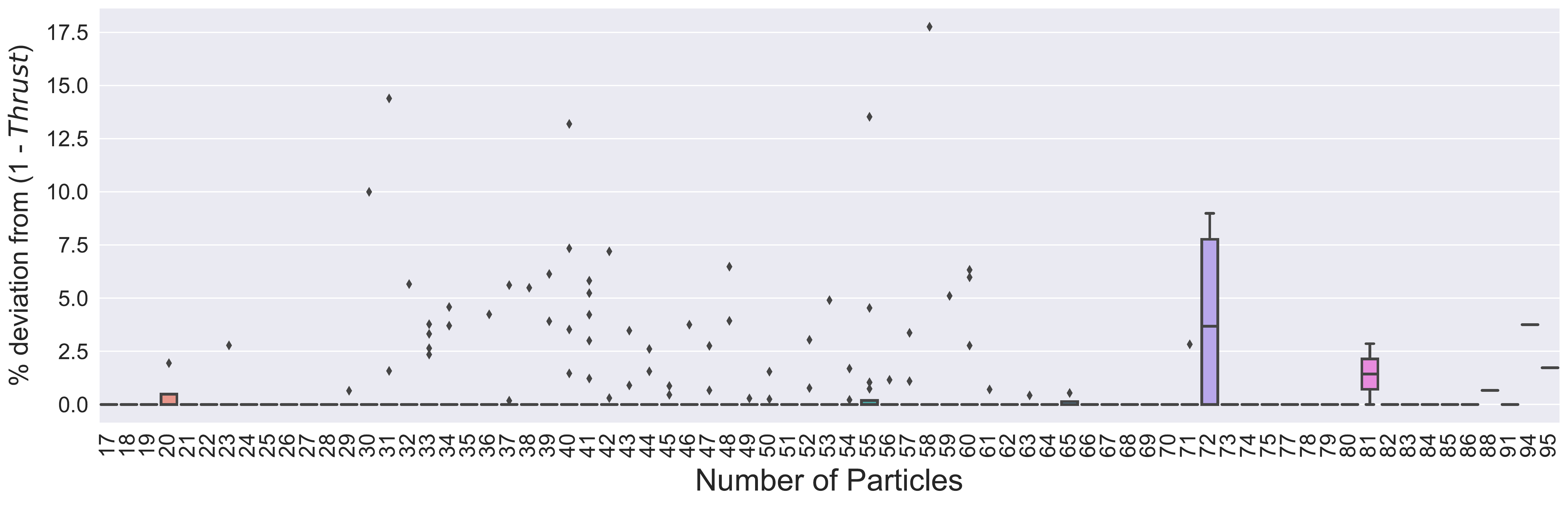}
\label{fig:per_saresults}
}
\caption{
Continuation of \Fig{per_devitation} for (d) the Advantage QPU with tuned setting, (e) using the forward annealing results without tuning as a seed for classical iterative improvement, and (f) classical simulated annealing.
}

\end{figure*}

\end{document}